\documentclass[11pt,english]{article}

\usepackage[T1]{fontenc}
\usepackage[latin1]{inputenc}
\usepackage{babel}
\begin{document}

\begin{center}
{\bf A  LOW MATTER DENSITY  DECAYING VACUUM COSMOLOGY FROM COMPLEX METRIC}
\par
\vspace{.15in}
Moncy V. John$^{\dag }$ and K. Babu Joseph,
\par
\medskip
Department of Physics,  Cochin University of Science and Technology,
Kochi, India.
\par

\vspace{1in}
{\bf Abstract}

\end{center}
A low matter density decaying 
vacuum cosmology is proposed on the assumption that the universe's radius is a 
complex quantity $\hat {R}$ if it is regarded as having a zero energy- momentum 
tensor. But we find that when the radius is real, it contains matter. Using the 
Einstein- Hilbert action principle,  the physical scale factor $ R(t) = \mid 
\hat {R}(t) \mid $ is obtained as equal to $ (R_{0}^{2} + t^{2} )^{1/2} $ with 
$R_{0}$ representing the finite radius of the universe at $t=0$. The resulting 
physical picture is roughly a theoretical justification of the old Ozer-Taha 
model. The new model is devoid of all cosmological problems. In particular, it 
confirms the bounds on $H_{P}$, the present value of 
Hubble parameter: $ 0.85 < H_{P} t_{P} < 1.91 $ and faces no age problem. 
We argue that the total 
energy density consists of parts corresponding to relativistic/ nonrelativistic 
matter, a positive vacuum energy, a negative energy and a form of matter with 
equation of state $ p_{K} = -\frac {1}{3} \rho _{K} $ (textures 
or generally K-matter), and the following predictions are made for the present 
nonrelativistic era: $
 \Omega _{M, \hbox {n.rel.}} \approx 2/3, \qquad 
 \Omega _{V. \hbox {n.rel.}} \approx 1/3, \qquad
 \Omega \_ \ll 1, \qquad \Omega _{K} \approx 1 $ where a parameter 
corresponding to K-matter is taken to be unity. It is shown that the space-time with complex 
metric has signature changing properties.  Using quantum cosmological 
considerations, it is shown that the wave function is peaked about the 
classical contour of evolution and the minimum radius $R_{0}$ of the nonsingular 
model is predicted as comparable with the Planck length.

---------------------------------------------------------------------------------

 $^{\dag}$Permanent Address : Department of Physics, St. Thomas College, 

Kozhencherri, Kerala, India 
689 641.
\vspace{2in}

\section{ Introduction}
The hot big bang model [1-3], which  is based 
upon the mathematically simple assumptions of spatial homogeneity and 
isotropy through the maximally symmetric Friedmann-Robertson-Walker (FRW) 
metric,  gives 
reliable and tested accounting of the history of the universe 
from very early times. But in this 
classical general relativistic model, a singularity (the big bang) in the 
past is  unavoidable and there are several 
`problems' associated with the model like the singularity \cite{narpad}, horizon \cite{horizon}, 
flatness \cite{dicke}, monopole \cite{brand}, cosmological constant \cite{weinpap} and age 
\cite{freedpi} problems.
Several attempts were made in the past to overcome these difficulties. 
It is hoped that the singularity problem may be obviated by quantum gravity 
effects which are expected to become important at the earliest epochs when 
temperature was as high as the Planck temperature \cite{halli}. Among other approaches, 
the most widely discussed idea is the occurrence of 
an inflationary scenario \cite{guth123} in the 
early universe, after the Planck epoch. Inflationary models envisage an era of 
exponential expansion, which is achieved by the action of an enormous 
vacuum energy, usually the potential energy of a weakly coupled scalar field 
initially displaced from the minimum of its potential. Entropy production is 
a crucial ingredient in this theory. It is found that inflation can drive 
the matter density to a value extremely close to the critical density 
so that universe appears flat from very early times. 
In a similar way, it solves 
the horizon and monopole problems. But the singularity and the cosmological 
constant problems are not  solved by this model alone. Some recent 
measurements \cite{freedpi} of the Hubble parameter pose another problem to the 
inflationary as well as standard (hot big bang) flat models without a cosmological constant. 
In these models, the combination 
$ H_{P}t_{P}$,  where $H_{P}$ is the present value of Hubble parameter 
and $t_{P}$ is the present age of the universe,  is predicted to have a value 
2/3. But the above observations restrict this value to be lying in a range 
$ 0.85 < H_{P} t_{P} < 1.91 $  . Introduction of a cosmological 
constant to save the inflationary models from this `age crisis'  
may aggravate the cosmological constant problem  
since one cannot distinguish vacuum energy from a cosmological 
constant \cite{kolbturner}. In addition, the several competing models in inflationary cosmology 
differ in the choice of scalar field potential as well as initial conditions 
which themselves use physics untested in the laboratory and remain largely
speculative and phenomenological.

However, it is widely believed that inflation is the unique 
mechanism capable of solving the horizon and flatness problems  and 
of explaining the formation of large scale structures.
It was recently argued \cite{hu} that an accelerating scale factor 
$R(t) $ and entropy production are essential conditions for a dynamical 
solution of  the horizon and flatness problems. Also it was pointed out \cite{liddle}
that the condition $\ddot {R} > 0 $ can produce the correct density 
perturbations required to explain structure formation in an 
inflationary evolution. But
it should be noted that their arguments do not require the presence of any 
scalar field or exponential expansion. In this context, it is desirable to 
 look for models  having $\ddot {R} > 0$ and 
entropy production, though without any exponential expansion. 
 Ozer and Taha 
have proposed such a model \cite{ozertaha} and  shown that 
the main cosmological problems including the singularity 
and cosmological constant problems can be solved by postulating that 
the equality of matter density and critical density is a time-independent 
feature (the critical density assumption - 
which helps them to circumvent the flatness problem ) 
and also that there is entropy 
production. They obtained a vacuum density which varies as $R^{-2}$. 
Such time-varying cosmological constant models were investigated by 
several authors [15-23]. 
Chen and Wu  suggested a model \cite{chen} based on the argument that some general 
dimensional considerations in line with quantum cosmology requires an $ R^{-2}$ 
dependence of the effective cosmological constant. This approach was developed by 
several authors [24-33].
 With the help of Landau-Lifshitz theory for 
nonequilibrium fluctuations, Pavon \cite{pavon} has analysed the models of  
Ozer-Taha, Chen-Wu, Freese $et.al $ \cite{freese} and Gasperini 
\cite{gasperini} and found that the 
former two models successfully pass their test of thermodynamic 
correctness.

In this paper, we present a nonsingular cosmological model  by making the 
assumptions that (i) the universe contains a zero energy-momentum tensor; 
(ii) it is  closed with a complex scale factor $\hat {R} (t)$ 
in the FRW metric  and also that (iii) it 
obeys the Einstein-Hilbert action principle. We argue that the physical 
scale factor is the real quantity $ R(t) = \mid \hat {R} (t) \mid $. 
With no further assumptions, we show that the evolution of R(t) is in such a way
that the closed real or observed universe with this real scale factor is 
nonempty and nonsingular and is an acceptable cosmological model.
 The evolution 
of R(t) coincides with that in the relativistic era 
of the model in \cite{ozertaha} and vacuum energy density varies as 
prescribed in \cite{chen}. The model is 
claimed to   have no cosmological problems mentioned above. While Abdel-Rahman 
\cite{abdel} has sought to generalise the Ozer-Taha model using the Chen-Wu 
prescription under the 
impression that the critical density assumption of the former  
 is hard to justify, our model derives both approaches.

The minimum value of the scale factor $R_{0}$ of this nonsingular 
model is not fixed by the classical theory. We have shown that a quantum 
cosmological treatment will help us to predict this value as comparable 
to the Planck length.

The paper is organised as follows. Section 2 summarises the relevant features of the 
 model in \cite{ozertaha}. In section 3, the gross features of the new model are 
introduced while in section 4, the energy-momentum tensor of the real 
universe and the solution of the cosmological problems are discussed. 
Section 5 deals with the quantisation of the system and section 6 comprises 
our conclusion.

This work is a significant improvement over that reported  in a recent letter \cite{jj}.
 
\section{ Ozer and Taha Model }

Ozer and Taha \cite{ozertaha} derived their model by starting from the field equations

\begin{equation}
R_{\mu \nu} - \frac {1}{2} R^{\sigma }_{\sigma} g_{\mu \nu} = 
-8\pi G \left[ T^{(M)}_{\mu \nu } + T^{(V)}_{\mu \nu} \right].
\end{equation}
where the total energy-momentum tensor of the universe is assumed to be the sum of 
 $ T^{(M)}_{\mu \nu}$, the energy-momentum tensor due to
  relativistic/ nonrelativistic matter, given by
 
 \begin{equation}
 T^{(M)}_{\mu \nu} = p_{M} g_{\mu \nu} + ( p_{M} + \rho _{M}) U_{\mu} U_{ \nu}
 \end{equation}
 and $T^{(V)}_{\mu \nu}$, which is the energy momentum tensor due to vacuum
 
 \begin{equation} 
 T^{(V)}_{\mu \nu} = -\rho_{V} g_{\mu \nu}
 \end{equation}
 Here $T^{(M)}_{\mu \nu}$ or $T^{(V)}_{\mu \nu}$ are not separately conserved.
 Only their sum is a conserved quantity. The above assumptions lead to 
 
 \begin{equation}
\left[ \frac {\dot {R}}{R}\right] ^{2} + \frac {k}{R^2} =\frac  {8\pi G}{3}(\rho _{M} + \rho _{V})
    \end{equation}
   and 

 \begin{equation}
   \frac {d(\rho _ {M}R^{3})}{dt} + p _ {M} \frac {dR^{3}} {dt} + R^{3} \frac {d\rho _ {V}}{dt} =0
   \end{equation}
   In \cite{ozertaha} it was noted that if there is to be entropy production, the condition
   $\frac { d\rho _ {V}}{dt} < 0$ should be imposed.
   Also they imposed the cri\-tical den\-sity as\-sumption
    $\rho _ {M} = \rho _ {C} \equiv \frac {3}{8\pi G} \left[ \frac {\dot {R}}{R}\right]^{2}$.
  With these assumptions, equations (4) and (5) give, for an expanding universe, 

   \begin{equation} 
   k=+1
   \end{equation}
   and
   
   \begin{equation}
   \rho _ {V} = \frac {3}{8\pi G} \frac {1}{R^{2}}
   \end{equation}
   In the relativistic era  where $p _ {M} = \frac{1}{3} \rho _ {M}$,
   the solutions are 
   \begin{equation}
   \rho _ {M} = \frac{3}{8\pi G} \left[\frac{1}{R^{2}} - \frac{R^{2} _ {0}}{R^{4}}\right]
   \end{equation}
   and
   
   \begin{equation}
   R^{2} (t) = R^{2} _ {0} + t^{2}
   \end{equation}
   \par
   The above evoluton of the scale factor helps them to solve all the main cosmological
   problems . However, this model does not explain the near
   equality of matter density and critical density in the present universe. 
Instead, this equality  is their ansatz. To describe the epochs after the 
relativistic one, they make several additional assumptions but these are of 
little interest in the present paper.
   
   \section{Derivation of the New Model}
\medskip 
 We start with the FRW metric  given by
\par
\begin{equation}
ds^{2} = dt^{2} - R^{2}(t)\left[\frac{dr^2}{1-kr^2} + r^{2} d\theta ^{2} + r^{2}\sin ^{2}\theta  d\phi ^{2}\right]
\end{equation}
If we make a substitution $R(t) \rightarrow \hat{R}(t) = R(t)e^{i\beta }$ 
in  Eq.(10),
then the space-time has Lorentzian signature (+ - - -) when $\beta  = \pm n \pi$,  $ ( n = 0,  1,  2, ..)$
and it has Riemannian signature (++++) when
$\beta  =\pm  (2n +1)\pi /2$, $( n = 0,  1,  2, ..)$. 
Such complex substitutions are familiar in relativity. For example, it is well known that open 
and closed FRW models, de Sitter and anti de Sitter space-times, Kerr and Schwarzchild metrics etc.
are related by  complex substitutions \cite{fla}.
Let  the  solutions $R(t)$  be   in   the     form
$R_{o} e^{\alpha (t)}$ . Then the above expression
 becomes $\hat{R}(t) = R_{o} e^{\alpha (t)+i\beta }$.
Interesting physics appears if we assume that the time-dependence of
the scale factor is shared by $\beta $ also; i.e., $\beta  = \beta (t)$, 
an assumption consistent with the homogeneity and isotropy conditions. 
Then the signature of the metric changes  when $ \beta $ varies from 
$0 \rightarrow \pi /2$ etc.
A signature change of the metric in the early universe is conceived 
and is a matter of hot debate in the current literature \cite{hh,sign}, though this 
appears in a different manner and context than ours.
Greensite \cite{sign} and Hayward \cite{sign} have even used a complex 
lapse and obtained signature changing features.
Our  ansatz   is   to replace $R(t)$ in equation (10) with
\par
\begin{equation}
\hat{R}(t) = R(t) e^{i\beta (t)} = R_{o} e^{\alpha (t)+i\beta (t)} \equiv  x(t) +i\; y(t)
\end{equation}
We further assume that this model of the universe with a complex scale factor 
is closed (ie., k = +1) and has a zero energy-momentum tensor.
The   evolution of $\hat{R}(t)$ is dictated by  the  
Einstein-Hilbert action principle, where the action is \cite{kolbturner} 
\par
\begin{equation}
S = \frac{-1}{16\pi G}\int d^{4}x (-g)^{1/2} {\cal R}
\end{equation}
Here ${\cal R}$   is the Ricci scalar 
\begin{equation}
{\cal R} = \frac{6}{N^{2}}\left[\frac{\dot{\hat{R}}}{\hat{R}}\right]^{2} -\frac{6}{\hat{R}^{2}}
\end{equation}
and $N$ is the lapse  function. Overdots  denote  time  derivatives.
 Using this and integrating the space part, we get
 
 \begin{equation}
 S = - \frac {3\pi}{4G} \int N \; dt \left[ \frac {\hat {R} \dot {\hat {R}}^{2}}{N^{2}} - \hat {R} \right]
 \end{equation}
We have included a boundary term to remove second derivatives in (14).
 Minimising this  action with respect to variations 
 of $\hat {R}$ and $N$ and fixing the gauge 
$ N = 1 $, we get the field equation 
\begin{equation}
2\frac {\ddot {\hat {R}}}{\hat {R}} + \left[ \frac {\dot {\hat {R}}}{\hat {R}}\right]^{2} + \frac {1}{\hat {R}^{2}} = 0
\end{equation}
and the constraint equation

\begin{equation}
\left[\frac { \dot {\hat {R}}}{\hat {R}} \right]^{2} + \frac {1}{\hat {R}^{2}} = 0.
\end{equation}
With $\hat {R}(t) = x(t) +i\; y(t) $ and $ x_{0}$, $ y_{0} $ constants, these 
equations may be solved to get

\begin{equation} 
\hat {R}(t) = x_{0} + i\; (y_{0} \pm t)
\end{equation}
We can choose the origin $ t= 0 $ 
such that $\hat {R} (0) = x_{0}$.
Relabelling $x_{0} \equiv R_{0}$
and assuming $R_{0} \neq 0$, we get,
\par
\begin{equation}
\hat {R} (t) = R_{0} \pm i\; t 
\end{equation}
This equation gives the contour of evolution of $\hat {R}(t)$ which is  
a straight line parallel to the imaginary axis. At $t=0$, this leaves the 
signature of space-time Lo\-re\-ntz\-ian but as $t \rightarrow \infty $ it becomes 
almost Riemannian. This need not create any 
conceptual problem since here we are considering
only an unperceived universe whose existence is our ansatz. 
(Simple physical intuition would give a signature `Riemannian at early times 
and Lorentzian at late' if it was for the physical universe we live in with 
matter contained in it. But in the above, we have a signature change in the 
opposite manner for the unphysical universe devoid of matter
and this need not contradict our physical intuition).
The connection with a closed real or physical universe is 
obtained by noting from the above that
\par
\begin{equation}
R^{2}(t) = \mid \hat {R} (t) \mid ^{2} = R^{2}_{0} + t^{2}
\end{equation}
This is the same equation (9) which governs the evolution of scale factor 
in the relativistic era of the Ozer-Taha model \cite{ozertaha}.

\section {The Real Universe}
 
From (11) and (18), we get

\begin{equation}
\beta (t) = tan ^{-1} ( \frac {\pm t} {R_{0}})
\end{equation}
and also

\begin{equation}
\dot {\beta }(t) = \frac {\pm R_{0}}{R^{2}(t)} = \frac {\pm cos ^{2} \beta }{R_{0}}
\end{equation}
With the help of these equations we observe that the real parts of the field 
equation (15) and the constraint equation (16) can be rewritten in terms 
of $R= R(t) = \mid \hat {R}(t) \mid $ as 

\begin{equation}
\frac {\ddot {R}}{R} = \frac {R^{2}_{0}}{R^{4}}, \qquad \left[ \frac 
{\dot {R}}{R} \right] ^{2} + \frac {1}{R ^{2}} = \frac {2}{R^{2}} - \frac 
{R^{2}_{0}}{R^{4}}  
\end{equation}
whose solution is $R(t) = (R_{0}^{2} + t^{2})^{1/2}$, as obtained in (19). 
We see that the real quantity $R(t)$ may be considered as the scale factor of 
a nonempty FRW universe. Equations (22) are appropriate for a closed FRW model
 with real 
scale factor R and with total energy density and total pressure  given by 
\begin{equation}
\tilde{\rho}  = \frac{3}{8\pi G}\left[\frac{2}{R^{2}} -\frac{ R^{2}_{0}}{R^{4}}\right]
\end{equation}
\begin{equation}
\tilde{p}   ={-\frac{1}{8\pi G}}\left[ \frac {2}{R^{2}} + 
\frac{R^{2}_{0}}{R^{4}} \right]
\end{equation}
respectively whose breakup can be performed in many ways. First let us assume, as
 done in \cite{ozertaha}, that 

\par
\begin{equation}
\tilde{\rho}  =  {\rho} _{M} +  {\rho} _{V}
\end{equation}
\begin{equation}
\tilde{p} =  {p}_{M} +  {p}_{V}
\end{equation}
where $M$ refers to matter and $V$ to  vacuum  terms.  We   also  write
  the relations between the pressure   due  to matter $ {p}_{M}$,
  that  due to  vacuum $ {p}_{V}$   and   the   corresponding  energy
densities ( equations of state) in the form
\par
\begin{equation}
 {p}_{M} = w\;  {\rho} _{M}
\end{equation}
\begin{equation}
 {p}_{V} = -  {\rho} _{V}
\end{equation}
where $w = 1/3$ for the relativistic era  and $w = 0$   for  the
nonrelativistic era. Solving (23)  and (24) using  (25)-(28),  one
gets
\par
\begin{equation}
 {\rho} _{M} = \frac{4}{8\pi G(1+w)}\left[\frac{1}{R^{2}} -\frac{R^{2}_{o}}{R^{4}}\right]
\end{equation}
\medskip
\begin{equation}
 {\rho _{V}}  = \frac{1}{8\pi G(1+w)}\left[ \frac {2(1+3w)}{R^{2}} + \frac {R^{2}_{0}(1-3w)}{R^{4}}\right]
\end{equation}
For  a relativistic matter dominated universe,  the  matter  density 
 $ {\rho} _{M,\hbox{rel.}} $ and the vacuum density ${\rho} _{V,\hbox{rel.}}$  are
\par
\begin{equation}
 {\rho }_{M,\hbox {rel.}}  = \frac {3}{8\pi G}\left[ \frac {1}{R^{2}} - \frac{R^{2}_{0}}{R^{4}}\right]
\end{equation}
\begin{equation}
 {\rho} _{V,\hbox{rel.}}  =\frac {3}{8\pi G}\frac {1}{R^{2}}
\end{equation}
From (22),  the critical density of the  real universe is
\par
\begin{equation}
\rho_{C}  \equiv \frac{3}{8\pi G} H^{2}  = \frac{3}{8\pi G}\left[\frac{1}{R^{2}} - \frac{R^{2}_{o}}{R^{4}}\right]
\end{equation}
where $ H $ , the value of the Hubble parameter 
is assumed to coincide with that predicted by the model. (We can see  that 
this is indeed the case in the present epoch by evaluating the combination 
$H_{P}t_{P}\equiv \left[ \frac {\dot {R}} {R}\right] _{P} t_{P}$ 
for $R_{0} \ll R_{P}$ which is found to be nearly equal to unity.) 
Then the ratios of  density  to  critical  density  for  matter  and
vacuum energy in the relativistic era are
\par
\begin{equation}
 {\Omega} _{M,\hbox {rel.}}\equiv \frac { {\rho }_{M,\hbox {rel.}}}{ {\rho }_{C}}=1
\end{equation}
\begin{equation}
 {\Omega}_{V,\hbox{rel.}}\equiv \frac{ {\rho}_{V,\hbox{rel.}}}{ {\rho}_C}\approx 1\qquad for\qquad R(t) \gg R_{0}
\end{equation}
For a universe dominated by nonrelativistic  matter,  the  condition
$w = 0$ may  be used in (29) and (30). In  this case,
\par
\begin{equation}
 {\Omega}_{M,\hbox{n.rel.}}=4/3
\end{equation}
\begin{equation}
 {\Omega} _{V,\hbox{n.rel.}} \approx  2/3 \qquad  for\qquad R(t) \gg R_{0}
\end{equation}
It  may be noted that (31) and (32) are the same  expressions
as those  obtained  in \cite{ozertaha} and (34) is their ansatz. 
But  the last two results  for  the
nonrelativistic era are outside the scope of that model.
\par
In the above, we have assumed $\tilde {\rho } = \rho _{M} + \rho _{V} $ 
following the example in \cite{ozertaha}. But this splitup is in no way 
unique. 
Equation (29) gives $\rho _{M} = 0$     at $t$ = 0.
In order to   avoid this less 
probable result, we  assume that 
the term $-\frac {3}{8\pi G} \frac {R_{0}^{2}}{R^{4}}  $ in $\tilde{\rho} $   
is  an  energy  density 
appropriate   for   negative   energy
relativistic  particles. The  pressure   $ {p}\_$    corresponding
to this negative energy density $ {\rho } \_ $  is also negative. Negative energy 
densities in the universe were postulated earlier \cite{hn}.
Such an assumption has the further advantage of making the expressions for 
$\rho _{M} $ and $\rho _{V}$ far more simple and of 
conforming  to the Chen  and Wu \cite{chen} prescription of  a pure $R^{-2}$
variation of vacuum density  (though, it should be noted that 
the Chen-Wu arguments, with $R_{0}$ identified as the Planck length, 
 are not against the form (30) for $\rho _{V}$ 
since the term which contains $\frac {R_{0}^{2}}{R^{4}}$ 
becomes negligibly small when compared to the $R^{-2}$ contribution 
within a few Planck times).   Thus we use a  modified
ansatz in this regard ( instead of (25)- (28)),
\par
\begin{equation}
\tilde{\rho}  =  {\rho} _{M} +  {\rho} _{V} +  {\rho} \_
\end{equation}
\begin{equation}
\tilde{p} =  {p}_{M} + { p}_{V} +  {p}\_
\end{equation}
\begin{equation}
{p}_{M} = w\;  {\rho} _{M}
\end{equation}
\begin{equation}
 {p}_{V} = -  {\rho} _{V}
\end{equation}
\begin{equation}
 {p}\_ = \frac{1}{3}  {\rho} \_
\end{equation}
and
\par
\begin{equation}
 {\rho} \_ = - \frac{3}{8\pi G}\frac{R^{2}_{o}}{R^{4}}
\end{equation}
and solving  (23) and  (24)  with   these   choices,   the results are,
\par
\begin{equation}
 {\rho}_{M}  =\frac{4}{8\pi G(1+w)}\frac{1}{R^{2}}
\end{equation}
\begin{equation}
 {\rho}_{V}  =\frac{2(1+3w)}{8\pi G(1+w)}\frac{1}{R^{2}}
\end{equation}
so that
\par
$$
 {\Omega} _{M,\hbox{rel.}} \approx  1 \qquad  {\Omega} _{V,\hbox{rel.}} \approx  1
$$
\begin{equation}
 {\Omega} _{M,\hbox{n.rel.}} \approx 4/3 \qquad  {\Omega} _{V,\hbox{n.rel.}} \approx  2/3
\end{equation}
$$
 {\Omega}\_\equiv\frac { {\rho}\_}{\rho_{C}} \ll 1 
$$
for $R(t) \gg R_{0} $. The predictions for $\Omega_{M}$ are marginal, though 
not ruled out by observations. Many authors [39-41]
seriously consider the existence 
of a new form of matter in the universe (called K-matter \cite{kolb} - 
perhaps a stable texture \cite{kam}) 
with the equation of state $ p_{K} = - \frac {1}{3} \rho _{K} $ and which 
decreases  as $R^{-2}$.  This leads to the idea of a low density closed 
universe \cite{kam}. If we accept this as probable,  the prediction 
 for $ \Omega_{M} $ will be well within 
the observed range of values. In this case we include a term $ \frac {3}{8\pi G}
\frac {K}{R^{2}}$ to the r.h.s of (38) so that
\begin{equation}
\rho _{M} = \frac {3}{8\pi G} \frac {2/3}{(1+w)} \frac {(2-K)}{R^{2}}
\end{equation}
and 

\begin{equation}
\rho _{V} = \frac {1}{8\pi G } \frac {(1+3w)}{(1+w)} \frac {(2-K)}{R^{2}}
\end{equation}
For a typical value $ K=1$ \cite{kolb}, the predictions for $R\gg R_{0}$ are
$$
\Omega _{M,\hbox {rel.}} \approx 1/2, \qquad \Omega _{V, \hbox {rel.}} \approx 1/2,
$$
\begin{equation}
 \Omega _{M, \hbox {n.rel.}} \approx 2/3, \qquad \Omega _{V. \hbox {n.rel.}}
\approx 1/3, 
\end{equation}
$$
 \Omega \_ \ll 1, \qquad \Omega _{K} \approx 1
$$
\medskip
It is easy to see that the conservation law for total energy
\par
\begin{equation}
\frac{d(R^{3}\tilde{\rho})}{dt} = -\tilde{p}\;\frac{dR^{3}}{dt}
\end{equation}
is  obeyed, irrespective of  the  ansatze   regarding   the
detailed structure of $\tilde{\rho} $.
\par
\medskip
Though the model makes clear cut predictions regarding the total energy 
density $\tilde {\rho}$ and total pressure $\tilde {p}$ as given by (23) and 
(24), the decomposition of these do not follow from any fundamental principles, 
except for those heuristic reasons we put forward. 
 But the solutions of cosmological problems are generic to the model. 
In \cite{jj} , we have noticed that if thermal radiation at  temperature $T$ 
is associated  with
energy densities ${\rho} _{M}$  and $ {\rho} \_$ (as given by (43) and (44))  according to the relation
\par
\begin{equation}
 {\rho}_{M}  +{\rho} \_ = \frac{1}{30} \pi ^{2} N(T) T^{4}
\end{equation}
(where $N(T)$ is the effective number
 of spin degrees  of  freedom   at
temperature $T$ and the units are such that $\hbar = c = k_{B} = 1$),
then the solution  of  cosmological  problems as given in \cite{ozertaha} can  be
performed  in  the present model also   without   any   alterations.

\par
 From (51), the temperature at time $t$ in the relativistic era is
\par
\begin{equation}
T ={ \left[\frac{3}{8\pi G}\; {\bf .} \; \frac {30}{\pi ^{2}N(T)}\right]}^{1/4} \left[\frac{t^{2}}{(R^{2}_{o}+t^{2})^{2}}\right]^{1/4}
\end{equation}
as in \cite{ozertaha}. At $t = 0,$ $ {\rho} _{M} +  {\rho} \_ = T = S = 0$ 
  where $S$  is   the
entropy. Then the temperature increases to a maximum value
\par
\begin{equation}
T_{max} ={\left[\frac{3}{8\pi G}\; {\bf .} \; \frac{15}{2\pi ^{2}N(T) R_{0}^{2}}\right]}^{1/4}
\end{equation}
 at $t = R_{0}$ and then decreases monotonically.
From a quantum cosmological treatment to follow in this paper, we 
obtain a value $ R_{0} = (\frac {2G}{3\pi })^{1/2}  \sim  l_{p}$ , the 
Planck length. With this value and assuming $N(T) \approx 100$ throughout the 
relativistic era, we get $T_{max} \sim .25 G^{-1/2} $, which is comparable 
with Planck energy. Note that here also, we are getting 
results which were anticipated in \cite{ozertaha}. 
  
Even if we omit $\rho \_ $ from (51) and write in a general way 

\begin{equation}
\frac {3}{8\pi G} \frac {\gamma }{R^{2}} = \frac {1}{30} \pi ^{2} N(T) T^{4}
\end{equation}
where $\gamma = 1 - \frac {K}{2} $ is a constant of the order of unity, all the main cosmological problems 
can be shown to disappear. In this case, the temperature

\begin{equation}
T = \left[ \frac {3}{8\pi G}\; {\bf .} \; \frac {30 \gamma } { \pi ^{2} N(T)} \right] ^{1/4} 
\left[ \frac {1} {R_{0}^{2} + t^{2}} \right] ^{1/4}
\end{equation}
is a maximum at $t=0$. If $R_{0} = (\frac {2G}{3\pi })^{1/2}$, then $ T(0) \sim .36 \gamma ^{1/4}G^{-1/2} $, which 
is comparable with the Planck energy and as $t \rightarrow \infty $, T decreases
monotonically, in contradistinction with the case considered above. The 
solution of the horizon problem depends only on the behaviour of the scale factor 
and proceeds as before. It can be seen that entropy is produced at the rate 

\begin{equation}
\frac {dS}{dt} = 4\pi ^{2} \frac {3 \gamma } {8\pi G} \left[  \frac {8\pi G }
{3} \; {\bf .} \; \frac {\pi^{2} N(T)}{30 \gamma } \right] ^{1/4} \frac {t}{(R_{0}^{2} + t^{2})^{1/4}}
\end{equation}
which enables the solution of cosmological problems.
  
Lastly the present monopole density predicted in this case can be seen to 
be 

\begin{equation}
n_{m} (t_{P}) \leq \left[ R_{P}\; sin h ^{-1} ( \frac {T^{2} (0)}{T^{2}_{c}}) \right] ^{-3}
\end{equation}
where $T_{c} \sim 10^{15} GeV $ is the grand unification phase transition 
temperature. This is very close to that estimated in \cite{ozertaha}, of the order of $ 10^{-120} GeV^{4} $ 
or less, which is negligibly smaller than the critical density. Thus the monopole
problem is solved in this case also.

\medskip
Irrespective of the case we are considering, the model is nonsingular and 
there is no singularity problem. The solution of the age problem is also 
generic to the model. It may be noted that the model correctly predicts 
the value of the combination $H_{P} t_{P} \approx 1 $. 
This places the present theory in a more advantageous position than
 the standard flat and the inflationary models with a zero cosmological constant
 where this value is predicted to be equal to 2/3, which is not 
in the range of recently observed values mentioned in the introduction.
 
\indent Another  interesting feature  is  that  since   the   expansion
process is reversible and the basic equations   are  time   reversal
invariant,  we can extrapolate to $t <0$. This  yields   an   earlier
contracting phase  for the universe. Such a phase  was  proposed  by
Lifshitz and Khalatnikov \cite{lif}. If there was such  an  initial  phase,
causality  could  have  established itself much earlier   than   the
time   predicted in \cite{ozertaha}.
\par
The   model predicts creation  of  matter  at   present
with a rate of creation per unit volume given by
\par
\begin{equation}
\frac{1}{R^{3}}\frac{d(R^{3} {\rho} _{M})}{dt} \mid_{P}\; =  {\rho} _{M,P}\;H _{P}
\end{equation}
where $ {\rho} _{M,P}$  is the present matter density. 
In arriving at this  result,   we   have
made  use of the assumption of a nonrelativistic  matter   dominated
universe. Note that the creation rate is
only  one-third of that in the steady-state cosmology \cite{narlikar}. Since   the
possibility of creation of matter or radiation at  the required rate
cannot be ruled out at the present level of observation \cite{lima},
 this does not pose any serious objection.

\section {Quantisation}

It should be noted that the classical theory discussed above does not fix 
the minimum radius $ R_{0} $ (though we have assumed $ R_{0} \neq 0 $). 
We now show that if we quantise the system described by the action (14), its 
value may be predicted. The momentum conjugate 
to $ \hat {R} $ is 

\begin{equation}
\Pi _{ \hat {R}} = \frac { \partial L }{ \partial \dot {\hat {R}}} 
=-\frac {3 \pi}{2G} \hat{R} \dot {\hat {R}}
\end{equation}
The canonical Hamiltonian is 

\begin{equation}
H = -\frac {G}{3 \pi} \frac {\Pi _{\hat {R}}^{2}}{\hat {R}}- 
\frac {3\pi }{4G} \hat {R}
\end{equation}
The constraint equation H = 0 has the corresponding Wheeler-Dewitt equation 

\begin{equation}
(H- \epsilon) \Psi (\hat {R})= 0
\end{equation}
where we have introduced an arbitrary real constant to take account of a 
possible energy renormalisation in passing from the classical 
constraint to its quantum operator form, as done by Hartle and Hawking in \cite{hh}.
Choosing the operator ordering for the sake of simplicity of  
the solution, we get,

\begin{equation}
\frac {d^{2} \Psi (\hat{R})}{d \hat {R}^{2}} - (m^{2} \hat {R}^{2} 
+2m \epsilon \hat {R} ) \Psi (\hat {R}) = 0
\end{equation}
where $ m = \frac {3\pi }{2G}$. Making  a substitution $ \hat {S}
= \sqrt {m} (\hat {R} + \frac{ \epsilon }{m}) $, this becomes,

\begin{equation}
\frac {d^{2} \Psi (\hat {S})}{d \hat {S} ^{2}} + (\frac {\epsilon ^{2}}{m} -  \hat {S}^{2}) \Psi (\hat {S}) = 0
\end{equation}
The wave equation has ground state harmonic oscillator type solution for 
$\epsilon  =\sqrt { m} $:

\begin{equation}
\Psi (\hat {R} ) = {\cal {N}}\; \; \hbox {exp } \left[ -\frac {m}{2}(\hat {R} + \frac {1}{\sqrt {m}})^{2}\right]
\end{equation}
This is nonnormalisable, but it is not normal in quantum cosmology to 
require that the wa\-ve\-fun\-ct\-ion should be normalised \cite{halli}.
Our choice is further justified by noting that the probability density

\begin{equation}
\Psi ^{\star }\Psi = {\cal {N}}^{2}\;\; \hbox {exp}\; (my^{2})\; \; \hbox {exp} \left[-m(x + \frac {1}{\sqrt {m}})^{2}\right]
\end{equation}
is sharply peaked about the classical contour  given by 
equation (18), which is a straight line parallel to the imaginary axis 
with $ x$ remaining a constant. We can
identify $R_{0}$ with the expectation value of $x$;

\begin{equation}
R_{0} \equiv <x> = -\frac {1}{\sqrt {m}} = - \sqrt {\frac {2G}{3\pi }}  
\end{equation}
so that $\mid R_{0} \mid \sim l_{P} $, which is the desired result.
The $e^{\frac {m}{2} y^{2}} $ part of the wavefunction is characteristic of a 
Riemannian space-time with signature (+ + + + ). This is precisely the 
feature we should expect to correspond to the imaginary part in the scale factor. 

\section{Summary and Discussion}
Based on the postulate that the universe contains a zero energy- momentum
tensor and is closed with a complex scale factor in the (signature changing) 
FRW metric, we get an acceptable FRW model for the real universe we live in 
with real scale factor $R(t) = \mid \hat {R}(t) \mid $ and having a nonzero 
 energy- momentum tensor. The model makes definite predictions regarding the 
 conserved total energy density of the universe. Those probable candidates 
which constitute this total energy are relativistic/ nonrelativistic matter, 
vacuum energy, negative energy and textures. In the present epoch, the negative 
energy density which scales as $ R^{-4} $ is vanishingly small whereas all 
other contributions (which vary as  $R^{-2}$) can be substantial and 
well within the observational bounds. The model 
predicts matter creation, but it may have little observable effects. The 
variation of vacuum energy coincides with a wide class of decaying vacuum 
cosmologies and satisfactorily explains the cosmological constant problem. 
The model has its nearest kinship with the Ozer-Taha model, the first among
this class. Recent speculations on the existence of matter with an equation 
of state $p_{K} = -\frac {1}{3} \rho _{K} $, like textures, may lead to a 
low density closed universe in the present model also. This model for the 
real universe is free from the singularity, flatness, horizon, monopole, 
cosmological constant and age problems.

We should keep in mind that the system described by the action (14) is the 
unperceived universe and it should not be confused with the model for the 
real physical world we live in. At t=0, the unphysical space-time has 
Lorentzian signature. But for large $t$, $ \hat{R} $ makes the signature 
of the metric almost Riemannian. All the other signature changing models \cite{hh,sign} 
try to describe the physical universe itself and they suggest that the real 
universe might have suffered a signature change from Riemannian to Lorentzian 
at a very early epoch in such a way that at present it has Lorentzian signature.
 But in our model signature change occured for the unperceived `hidden' 
universe and that too from Lorentzian to almost Riemannian. The physical, 
perceivable universe is the usual FRW one with scale factor $R(t)$ and we have 
only established a correspondence with the former by arguing that the physical 
 scale factor is the modulus of $\hat {R}$. If we quantise the model for 
 the real universe by starting from equation (22), it would be one of 
 those cases contained in Fil'chenkov \cite{fil}. But we regard  the 
former space-time with the complex metric as the model for an underlying objective reality and hence we attempt 
to quantise this system. A remarkable result obtained on quantisation is that 
the simplest minimum energy wave function is sharply peaked about the classical contour of 
evolution of $\hat {R}$, just like the ground state harmonic oscillator wave 
function in quantum mechanics is peaked about the classical position of the 
particle. But we welcome the important difference with this analogy; ie., the 
quantum mechanical system in our case is not localised. In fact,
the wave function is not normalisable along the imaginary 
axis. If it was with real scale factor, the exponential growth of the 
wave function would correspond to some classically forbidden region, but 
in this case, we have the nonnormalisable part for the wave function along 
the imaginary axis; this result is just what we should expect 
since it corresponds to our classical system and cannot 
be termed as `classically forbidden'. Since we are confined to 
the minimum energy eigen state to satisfy the 
constraint equation, we cannot hope to obtain the classical trajectory 
$\hat {R}(t) = R_{0} \pm i \; t$. The most significant fact is that the 
quantum cosmological treatment helps us to predict the value of $R_{0}$, 
the minimum radius in the nonsingular model.

The expansion rate equation (19) is the coasting solution. The standard 
case of a coasting evolution is the Milne universe, with $\tilde {\rho } = 
\tilde {p} = 0 $. Ellis et. al. \cite{ellis} have obtained coasting solution 
for the universe with a scalar field under suitable potentials. Coasting evolution 
is obtained in theories with cosmic strings or textures as mentioned 
earlier. Since the expansion rate is faster than that in the hot big bang 
cosmology, it is suspected to interfere with the nucleosynthesis of the early 
universe. But that nucleosynthesis is not always a fatal impediment in 
a coasting cosmology is shown by Batra and Lohiya \cite{ellis}. The nucleosynthesis 
in the present model is yet to be worked out. When it comes to the cosmic microwave 
background radiation (CMBR), we note that the correct shape of its spectral 
distribution in the present model depends on the details of matter production. 
Lima \cite{lima} has shown that for cosmologies with photon creation, a 
new Planckian distribution can be obtained which is compatible with the 
present spectral shape of CMBR. Structure formation can be addressed in 
the theory by considering perturbations in the complex metric. This leads to the 
question as to what the real metric is in this situation. We believe that the 
scale factor of the real universe will continue to be the modulus of the 
scale factor in the complex metric and that it will give the desired results. 
However, this problem is also not considered in any detail here.
 We propose to discuss these issues, viz., 
 nucleosynthesis, CMBR and structure formation in future publications.
 It is also interesting to note that the evolution of 
the scale factor obtained in our model is analogous to that  of the conformal factor 
obtained by Padmanabhan \cite{pad} in the study of quantum stationary geometries 
and the connection is  to be explored.

Whereas the most well studied cosmological models like the hot big bang and
steady state models are phenomenological with regard to the
ene\-rgy\--mom\-entum tensor, the present model makes definite prediction of
this value.  Inflationary models enjoy wide popularity among physicists mainly
due to the exciting possibility that the standard model physics describing
fundamental particles at the microlevel join hands with the most macroscopic
theory conceivable, which is cosmology.  But though being invented to cure the
pathos of the hot big bang model, these models themselves find many of those
problems unsolvable.  But our model successfully handles all those issues which
are problematic in the big bang model.  The model is claimed to be derived at a
fundamental level.  Also it shows glimpses of the much sought after quantum
cosmological ideals.  As stated often, physics progresses through critical
reexamination of the fundamental issues and by unmasking new insights.  In this
context, we believe that the model present in this paper is worth pursuing.

\medskip
\noindent {\bf Acknowledgements}

We thank Prof. J. V. Narlikar and Prof. T. Padmanabhan 
for valuable comments and Prof. N. Dadhich  for pointing out the probable 
existence of K-matter to us. M.V.J. gratefully acknowledges the hospitality of 
the Inter University Centre for Astronomy and Astrophysics (IUCAA), Pune.

\end{document}